\documentclass[rnote]{aa} 

\usepackage{graphicx,url,twoopt,natbib}
\usepackage[varg]{txfonts}
\usepackage[pdftitle={Transits of extrasolar moons around luminous giant planets}, colorlinks = true, breaklinks = true, citecolor = blue, linkcolor = blue, urlcolor = blue, pdfauthor = {Ren\'{e} Heller}]{hyperref}
\usepackage{breakurl}
\bibpunct{(}{)}{;}{a}{}{,}    

\begin{document} 

\title{Transits of extrasolar moons around luminous giant planets}

\author{R.~Heller\inst{1}
           }

\institute{Max Planck Institute for Solar System Research, Justus-von-Liebig-Weg 3, 37077 G\"ottingen, Germany; \href{mailto:heller@mps.mpg.de}{heller@mps.mpg.de}
              }

\date{Received October 2, 2015; Accepted February 1, 2016}
 

\abstract{Beyond Earth-like planets, moons can be habitable, too. No exomoons have been securely detected, but they could be extremely abundant. Young Jovian planets can be as hot as late M stars, with effective temperatures of up to 2000\,K. Transits of their moons might be detectable in their infrared photometric light curves if the planets are sufficiently separated ($\gtrsim10$\,AU) from the stars to be directly imaged. The moons will be heated by radiation from their young planets and potentially by tidal friction. Although stellar illumination will be weak beyond 5\,AU, these alternative energy sources could liquify surface water on exomoons for hundreds of Myr. A Mars-mass H$_2$O-rich moon around $\beta$\,Pic\,b would have a transit depth of $1.5\times10^{-3}$, in reach of near-future technology.}


\maketitle

\section{Introduction}

\noindent
Since the discovery of a planet transiting its host star \citep{2000ApJ...529L..45C}, thousands of exoplanets and candidates have been detected, mostly by NASA's \textit{Kepler} space telescope \citep{2014ApJ...784...45R}. Planets are natural places to look for extrasolar life, but moons around exoplanets are now coming more and more into focus as potential habitats \citep{1987AdSpR...7..125R,1997Natur.385..234W,2006ApJ...648.1196S,2014AsBio..14..798H,2014OLEB...44..239L}. Key challenges in determining whether an exomoon is habitable or even inhabited are in the extreme observational accuracies required for both a detection, e.g., via planetary transit timing variations plus transit duration variations \citep{1999A&AS..134..553S,2009MNRAS.392..181K}, and follow-up characterization, e.g., by transit spectroscopy \citep{2010ApJ...712L.125K} or infrared (IR) spectral analyses of resolved planet-moon systems \citep{2014ApJ...796L...1H,2015ApJ...812....5A}. New extremely large ground-based telescopes with unprecedented IR capacities (\textit{GTM}, 1st light 2021; \textit{TMT}, 1st light 2022; \textit{E-ELT}, 1st light 2024) could achieve data qualities required for exomoon detections \citep{2015IJAsB..14..279Q}.

Here I investigate the possibility of detecting exomoons transiting their young, luminous host planets. These planets need to be sufficiently far away from their stars ($\gtrsim10$\,AU) to be directly imaged. About two dozen of them have been discovered around any stellar spectral type from A to M stars, most of them at tens or hundreds of AU from their star. The young super-Jovian planet $\beta$\,Pic\,b \citep[$11\pm5$ times as massive as Jupiter,][]{2014Natur.509...63S} serves as a benchmark for these considerations. Its effective temperature is about 1\,700\,K \citep{2014IAUS..299..277B}, and contamination from its host star is sufficiently low to allow for direct IR spectroscopy with \textit{CRIRES} at the \textit{VLT} \citep{2014Natur.509...63S}. Exomoons as heavy as a few Mars masses have been predicted to form around such super-Jovian planets \citep{2006Natur.441..834C,2014AsBio..14..798H}, and they might be as large as 0.7 Earth radii for wet/icy composition \citep{2015ApJ...806..181H}. The transit depth of such a moon ($10^{-3}$) would be more than an order of magnitude larger than that of an Earth-sized planet around a Sun-like star. Photometric accuracies down to 1\,\% have now been achieved in the IR using \textit{HST} \citep{2015arXiv151202706Z}. Hence, transits of large exomoons around young giant planets are a compelling new possibility for detecting extrasolar moons.


\section{Star-planet versus planet-moon transits}

\subsection{Effects of planet and moon formation on transits}

\noindent
Planets and moons form on different spatiotemporal scales. We thus expect that geometric transit probabilities, transit frequencies, and transit depths differ between planets (transiting stars) and moons (transiting planets). The H$_2$O ice line, beyond which runaway accretion triggers formation of giant planets in the protoplanetary disk \citep{1987Icar...69..249L,2007ApJ...664L..55K}, was at about 2.7\,AU from the Sun during the formation of the local giant planets \citep{1981PThPS..70...35H}. In comparison, the circum-Jovian H$_2$O ice line, beyond which some of the most massive moons in the solar system formed, was anywhere between the orbits of rocky Europa and icy Ganymede \citep{1974Icar...21..248P} at 10 and 15 Jupiter radii ($R_{\rm J}$), respectively. The ice line radius ($r_{\rm ice}$), which is normalized to the physical radius of the host object ($R$), was $2.7\,{\rm AU}/12.5\,R_{\rm J}\approx480$ times larger in the solar accretion disk than in the Jovian disk. With $r_{\rm ice}$ depending on $R$ and the effective temperature ($T_{\rm eff}$) of the host object as per

\begin{equation}\label{eq:r_ice}
r_{\rm ice} \propto \sqrt{ R^2 \, T_{\rm eff}^4 } \ \ ,
\end{equation}

\noindent
we understand this relation by comparing the properties of a young Sun-like star ($R=R_\odot$ [solar radius], $T_{\rm eff}=5000$\,K) with those of a young, Jupiter-like planet ($R=R_{\rm J}$, $T_{\rm eff}=1000$\,K). This suggests a $\sqrt{10^2 \times 5^4} = 250$ times wider H$_2$O ice line for the star, neglecting the complex opacity variations in both circumstellar and circumplanetary disks \citep{2015A&A...575A..28B}.

The mean geometric transit probabilities ($\bar{P}$) of the most massive moons should thus be larger than $\bar{P}$ of the most massive planets. Because of their shorter orbital periods, planet-moon transits of big moons should also occur more often than stellar transits of giant planets. In other words, big moons should exhibit higher transit frequencies ($\bar{f}$) around planets than giant planets around stars, on average. However, Eq.~(\ref{eq:r_ice}) does not give us a direct clue as to the mean relative transit depths ($\bar{D}$).

\subsection{Geometric transit probabilities, transit frequencies, and transit depths}
\label{sub:transits}

I exclude moons around the local terrestrial planets (most notably the Earth's moon) and focus on large moons around the solar system giant planets to construct an empirical moon sample representative of moons forming in the accretion disks around giant planets. This family of natural satellites has been suggested to follow a universal formation law \citep{2006Natur.441..834C}. We need to keep in mind, though, that these planets orbit the outer regions of the solar system, where stellar illumination is negligible for moon formation \citep{2015A&A...578A..19H}. However, many giant exoplanets are found in extremely short-period orbits (the ``hot Jupiters''); and stellar radial velocity (RV) measurements suggest that giant planets around Sun-like stars can migrate to 1\,AU, where we observe them today. Moreover, at least one large moon, Triton, has probably been formed through a capture rather than in-situ accretion \citep{2006Natur.441..192A}.

For the planet sample, I first use all RV planets confirmed as of the day of writing. I exclude transiting exoplanets from my analysis as these objects are subject to detection biases. RV observations are also heavily biased \citep{2004MNRAS.354.1165C}, but we know that they are most sensitive to close-in planets because of the decreasing RV amplitude and longer orbital periods in wider orbits. Thus, planets in wide orbits are statistically underrepresented in the RV planet sample, but these planets are not equally underrepresented as planets in transit surveys. Transit surveys also prefer close-in planets, as the photometric signal-to-noise ratio scales as ${\propto}~n_{\rm tr}$ \citep[$n_{\rm tr}$ the number of transits,][]{2012ApJS..201...15H}.

Mean values of $\bar{P}$ (similarly of $\bar{f}$ and $\bar{D}$) are calculated as $\bar{P}=(\sum_i^{N_{\rm p}}P_i)/N_{\rm p}$, where $P_i$ is the geometric transit probability of each individual RV planet and $N_{\rm p}$ is the total number of RV planets. Standard deviations are measured by identifying two bins around $\bar{P}$: one in negative and one in positive direction, each of which contains $\frac{1}{2} 68\,\%=34\,\%$ of all the planets in the distribution. The widths of these two bins are equivalent to asymmetric $1\,\sigma$ intervals of a skewed normal distribution.

The geometric transit probability $P = R/a$, with $a$ as the orbital semimajor axis of the companion. For the RV planets, $\bar{P}=0.028\,(+0.026, - 0.019)$.\footnote{Information for both $R$ and $a$ was given for 398 RV planets listed on\, \url{http://www.exoplanet.eu} as of 1 October 2015.} The $\bar{P}$ value of an unbiased exoplanet population would be smaller as it would contain more long-period planets. On the contrary, no additional detection of a large solar system moon is expected, and $\bar{P}=0.114\,(+0.080, - 0.045)$, for the 20 largest moons of the solar system giant planets, likely reflects their formation scenarios.

The transit frequency $f \approx 1/(2\pi) \sqrt{(GM)/a^3}$, where $G$ is Newton's gravitational constant and $M$ the mass of the central object, assuming that $M$ is much larger than the mass of the companion. For the RV planets, $\bar{f}=0.040\,(+0.052, - 0.037)$/day.\footnote{The orbital period was given for 612 RV planets} The value of $\bar{f}$, which would be corrected for detection biases, would be smaller with long-period planets having lower frequencies. For comparison, $\bar{f}$ of the 20 largest moons of the solar system giant planets is $0.170\,(+0.111, - 0.030)$/day.

Stellar limb darkening, star spots, partial transits, etc. aside, the maximum transit depth depends only on $R$ and on the radius of the transiting object ($r$) as per $D = \left(r/R \right)^2$. In my calculations of $D$, I resort to the transiting exoplanet data because $r$ is not known for most RV planets. For the transiting planets, $\bar{D}=3.00\,(+4.69, -2.40)\times10^{-3}$.\footnote{Information for both $r$ and $R$ was given for 1202 transiting planets.} It is not clear whether a debiased $\bar{D}$ value would be larger or smaller than that. This depends on whether long-period planets usually have larger or smaller radii than those used for this analysis. For comparison, $\bar{D}$ of the 20 largest moons of the solar system giant planets is $6.319\,(+3.416, -4.543)\times10^{-4}$.

\begin{figure}[t]
  \centering
  \includegraphics[width=.932\linewidth]{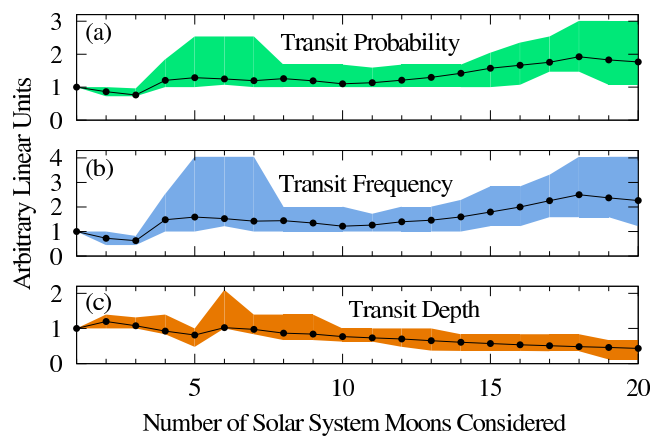}
  \caption{Dependence of (a) the mean geometric transit probability, (b) mean transit frequency, and (c) mean transit depth of the largest solar system moons on the number of moons considered. The abscissa refers to the ranking ($N_{\rm s}$) of the moon among the largest solar system moons.}
  \label{fig:cummulative}
\end{figure}

\begin{figure*}[t]
  \centering
  \includegraphics[width=0.477\linewidth]{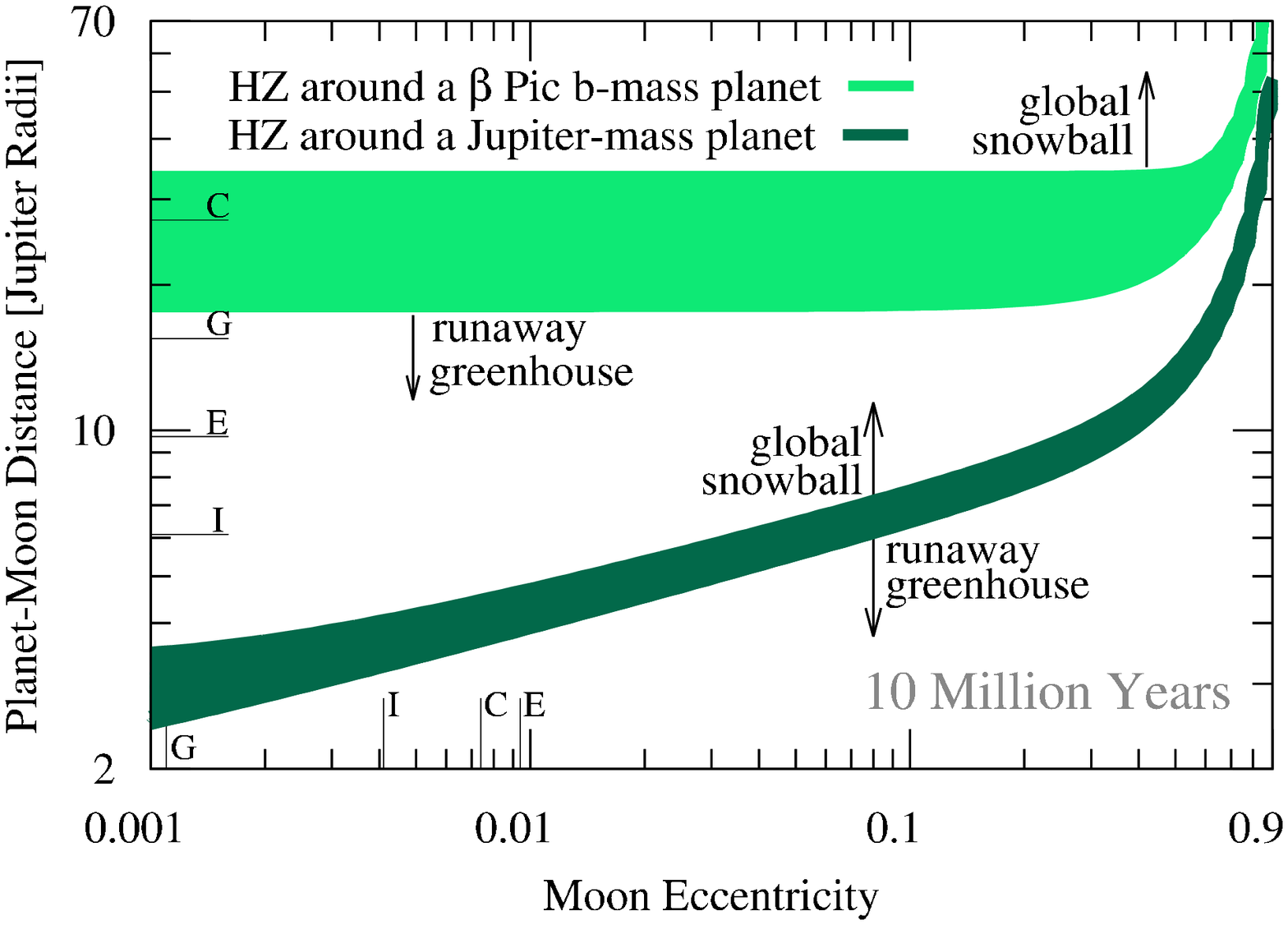}
  \hspace{.8cm}
  \includegraphics[width=0.447\linewidth]{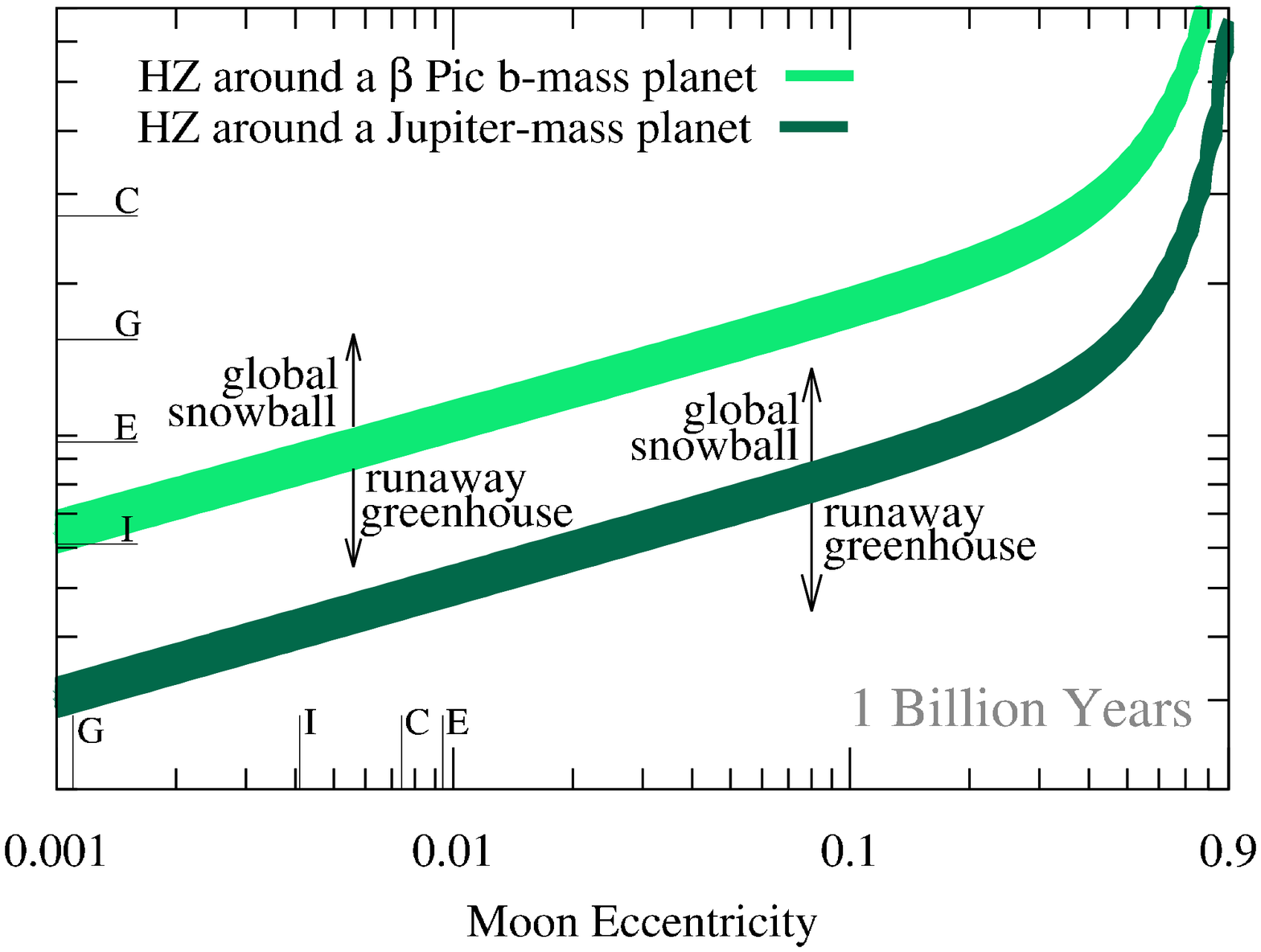}  
  \caption{Habitable zone (HZ) of a Mars-sized moon around a Jupiter-mass (light green) and a $\beta$\,Pic\,b-like $11\,M_{\rm J}$-mass (dark green) planet at $\gtrsim5$\,AU from a Sun-like star. Inside the green areas, tidal heating plus planetary illumination can liquify water on the moon surface. Values of the Galilean moons are denoted by their initials. \textit{Left:} Both HZs refer to a planet at an age of 10\,Myr, where planetary illumination is significant. \textit{Right:} Both HZs refer to a planet at an age of 1\,Gyr, where planetary illumination is weak and tidal heating is the dominant energy source on the moon.}
  \label{fig:exomoonHZ}
\end{figure*}

The choice of the 20 largest local moons for comparison with the exoplanet data is arbitrary and motivated by the number of digiti manus. Figure~\ref{fig:cummulative}(a)-(c) shows $\bar{P}$, $\bar{f}$, and $\bar{D}$ as functions of the number of natural satellites ($N_{\rm s}$) taken into account. Solid lines denote mean values, shaded fields standard deviations. The first moon is the largest moon, Ganymede, the second moon Titan, etc., and the 20th moon Hyperion. The trends toward higher transit probabilities (a), higher transit frequencies (b) and lower transit depths (c) are due to the increasing amount of smaller moons in short-period orbits. The negative slope at moons 19 and 20 in (a) and (b) is due to Nereid and Hyperion, which are in wide orbits around Neptune and Saturn, respectively.

The key message of this plot is that $\bar{P}(N_{\rm s}=20)$, $\bar{f}(N_{\rm s}=20)$, and $\bar{D}(N_{\rm s}=20)$ used above for comparison with the exoplanet data serve as adequate approximations for any sample of solar system moons that would have been smaller because variations in those quantities are limited to a factor of a few.

\subsection{Evolution of exomoon habitability}

An exomoon hosting planet needs to be sufficiently far from its star to enable direct imaging and to reduce contamination from IR stellar reflection. This planet needs to orbit the star beyond several AU, where alternative energy sources are required on its putative moons to keep their surfaces habitable, i.e., to prevent freezing of H$_2$O. This heat could be generated by (1) tides \citep{1987AdSpR...7..125R,2006ApJ...648.1196S,2009ApJ...704.1341C,2013AsBio..13...18H}, (2) planetary illumination \citep{2015IJAsB..14..335H}, (3) release of primordial heat from the moon's accretion \citep{1987Icar...69...91K}, or (4) radiogenic decay in its mantle and/or core \citep{1988Icar...76..437M}. All of these sources tend to subside on a Myr timescale, but (1) and (2) can compete with stellar illumination over hundreds of Myr in extreme, yet plausible, cases. (3) and (4) usually contribute $\ll1\,{\rm W\,m}^{-2}$ at the surface even in very early stages. Earth's globally averaged internal heat flux, for example, is $86\,{\rm mW\,m}^{-2}$ \citep{2007SSRv..129...35Z}, which is mostly fed by radiogenic decay in the Earth's interior.

The globally averaged absorption of sunlight on Earth is $239\,{\rm W\,m}^{-2}$. If an Earth-like object (planet or moon) absorbs more than $295\,{\rm W\,m}^{-2}$, it will enter a runaway greenhouse effect \citep{1988Icar...74..472K}. In this state, the atmosphere is opaque in the IR because of its high water vapor content and high IR opacity. The surface of the object heats up beyond 1\,000\,K until the globe starts to radiate in the visible. Water cannot be liquid under these conditions and the object is uninhabitable by definition. For a Mars-sized moon, which I consider as a reference case, the runaway greenhouse limit is at $266\,{\rm W\,m}^{-2}$, following a semianalytic model \citep[Eq.~4.94 in][]{2010ppc..book.....P}. If the combined illumination plus tidal heating (or any alternative energy source) exceed this limit, then the moon is uninhabitable.

On the other hand, there is a minim energy flux for a moon to prevent a global snowball state that is estimated to be 0.35 times the solar illumination absorbed by Earth \citep[$83\,{\rm W\,m}^{-2}$,][]{2013ApJ...765..131K}, which is only weakly dependent on the object's mass. I use the terms global snowball and runaway greenhouse limits to identify orbits in which a Mars-sized moon would be habitable, which is similar to an approach presented by \citet{2014AsBio..14...50H}. A snowball moon might still have subsurface oceans, but because of the challenges of even detecting life on the surface of an exomoon, I here neglect subsurface habitability. I calculate the total energy flux by adding the stellar visual illumination and the planetary thermal IR radiation absorbed by the moon to the tidal heating within the moon. Illumination is computed as by \citet{2015IJAsB..14..335H}, and I neglect both stellar reflected light from the planet and the release of primordial and radiogenic heat. Assuming an albedo of 0.3, similar to the Martian and terrestrial values, the absorbed stellar illumination by the moon at 5.2\,AU from a Sun-like star is $9\,{\rm W\,m}^{-2}$. I consider a Jupiter-mass planet and a $\beta$\,Pic\,b-like $11\,M_{\rm J}$-mass planet, both in two states of planetary evolution; one, in which the system is 10\,Myr old (similar to $\beta$\,Pic\,b) and the planet is still very hot and inflated; and one, in which the system has evolved to an age of 1\,Gyr and the planet is hardly releasing any thermal heat.

Planetary luminosities are taken from evolution models \citep{2013A&A...558A.113M}, which suggest that the Jupiter-mass planet evolves from $R=1.28\,R_{\rm J}$ and $T_{\rm eff}=536$\,K to $R=1.03\,R_{\rm J}$ and $T_{\rm eff}=162$\,K over the said period. For the young $11\,M_{\rm J}$-mass planet, I take $R=1.65\,R_{\rm J}$ \citep{2014Natur.509...63S} and $T_{\rm eff}=1\,700$\,K \citep{2014IAUS..299..277B}. For the 1\,Gyr version, I resort to the \citet{2013A&A...558A.113M} tracks, predicting $R=1.09\,R_{\rm J}$ and $T_{\rm eff}=480$\,K. Tidal heating is computed as by \citet{2013AsBio..13...18H}, following an earlier work on tidal theory by \citet{1981A&A....99..126H}.

Figure~\ref{fig:exomoonHZ} shows the HZ for exomoons around a Jupiter-like (dark green areas) and a $\beta$\,Pic\,b-like planet (light green areas) at ages of 10\,Myr (left) and 1\,Gyr (right). Planet-moon orbital eccentricities ($e$, abscissa) are plotted versus planet-moon distances (ordinate). In both panels, the HZ around the more massive planet is farther out for any given $e$, which is mostly due to the strong dependence of tidal heating on $M_{\rm p}$. Most notably, the planetary luminosity evolution is almost negligible in the HZ around the $1\,M_{\rm J}$ planet. In both panels, the corresponding dark green strip has a width of just $1\,R_{\rm J}$, details depending on $e$.

The HZ around the young $\beta$\,Pic\,b, on the other hand, spans from $18\,R_{\rm J}$ to $33\,R_{\rm J}$ for $e\lesssim0.1$ (left panel). This wide range is due to the large amount of absorbed planetary thermal illumination, which is the main heat source on the moon in these early stages. Planetary radiation scales as $\propto~a^{-2}$, causing a much smoother transition from a runaway greenhouse (at $18\,R_{\rm J}$) to a snowball state (at $33\,R_{\rm J}$) than on a moon that is fed by tidal heating alone, which scales as $\propto~a^{-9}$. In the right panel, planetary illumination has vanished and tides have become the principal energy source around an evolved $\beta$\,Pic\,b. However, the HZ around the evolved $\beta$\,Pic\,b is still a few times wider (for any given $e$) than that around a Jupiter-like planet; note the logarithmic scale.

\section{Discussion}

A major challenge for exomoon transit observations around luminous giant planets is in the required photometric accuracy. Although \textit{E-ELT} will have a collecting area 1\,600 times the size of \textit{Kepler}'s, it will have to deal with scintillation. The IR flux of an extrasolar Jupiter-sized planet is intrinsically low and comes with substantial white and red noise components. Photometric exomoon detections will thus depend on whether \textit{E-ELT} can achieve photometric accuracies of $10^{-3}$ with exposures of a few minutes.\footnote{Transits of a moon that formed at the H$_2$O ice line around a $\beta$\,Pic\,b-like planet \citep[$20\,R_{\rm J}$,][]{2015A&A...578A..19H} take about 1\,hr:13\,min.} Adaptive optics and the availability of a close-by reference object, e.g., a low-mass star with an apparent IR brightness similar to the target, will be essential. Alternatively, systems with multiple directly imaged giant planets would provide particularly advantageous opportunities, both in terms of reliable flux calibrations and increased transit detection probabilities. Systems akin to the HR\,8799 four-planet system \citep{2008Sci...322.1348M,2010Natur.468.1080M} will be ideal targets.

Transiting moons could also impose RV anomalies on the planetary IR spectrum, known as the Rossiter-McLaughlin (RM) effect \citep{1924ApJ....60...22M,1924ApJ....60...15R}. The RM reveals the sky-projected angle between the orbital plane of transiting object and the rotational axis of its host, which has now been measured for 87 extrasolar transiting planets\footnote{Holt-Rossiter-McLaughlin Encyclopaedia at \url{http://www2.mps.mpg.de/homes/heller}}. \textit{E-ELT} might be capable of RM measurements for large exomoons transiting giant exoplanets that can be directly imaged \citep{2014ApJ...796L...1H}.

Even nontransiting exomoons might be detectable in the IR RV data of young giant planets. The estimated RV 1$\sigma$ confidence achievable on a $\beta$\,Pic\,b-like giant exoplanet with a high-resolution ($\lambda/\Delta\lambda\approx100\,000$, $\lambda$ being the wavelength of the observed light) near-IR spectrograph mounted to the \textit{E-ELT} could be as low as $\approx70\,{\rm m\,s}^{-1}$ in reasonable cases \citep{2014ApJ...796L...1H}. The RV amplitude of an Earth-mass exomoon in a Europa-wide ($a\approx10\,R_{\rm J}$) orbit around a Jupiter-mass planet would be $43\,{\rm m\,s}^{-1}$ with a period of 3.7\,d. RV detections of super-massive moons, if they exist, would thus barely be possible even with \textit{E-ELT} IR spectroscopy. We should nevertheless recall that ``hot Jupiters'' had not been predicted prior to their detection \citep{1995Natur.378..355M}. In analogy, observational constraints could still allow for detections of a so-far unpredicted class of hot super-Ganymedes; they might indeed be hot due to enhanced tidal heating \citep{2013ApJ...769...98P} and/or illumination from the young planet \citep{2015IJAsB..14..335H}.

Moon eccentricities tend to be tidally eroded to zero in a few Myr. Perturbations from other moons or from the star can maintain $e>0$ for hundreds of Myr. With $e$ varying in time, exomoon habitability could be episodical. Uncertainties in the tidal quality factor ($Q$) and the 2nd order Love number ($k_2$), which scale the tidal heating as per ${\propto}k_2/Q$, can be up to an order of magnitude. However, because of the strong dependence on $a$, the HZ limits in Fig.~\ref{fig:exomoonHZ} would be affected by $<1\,R_{\rm J}$.

\section{Conclusions}

Large moons around the local giant planets transit their planets much more likely from a randomly chosen geometrical perspective and significantly more often than the RV exoplanets transit their stars. This is likely a fingerprint of planet and moon formation acting on different spatiotemporal scales. If the occurrence rate of planets per star is similar to the occurrence rate of large moons around giant planets (1-10 per system), the probability of observing a moon transiting a randomly chosen giant planet is at least four times higher than the probability of observing a planet transiting a randomly chosen star; planet-moon transits are at least four times more frequent than star-planets transits; the average transit depth of the transiting exoplanets is five times larger than the average transit depth of the twenty largest moons around the local giant planets. However, each solar system giant planet has at least one moon with a transit depth of $10^{-3}$. Following the gas-starved accretion disk model for moon formation \citep{2006Natur.441..834C,2015ApJ...806..181H}, a super-Ganymede around the young super-Jovian exoplanet $\beta$\,Pic\,b could have a transit depth of $\approx1.5\times10^{-3}$, which is 18 times as deep as the transit of the Earth around the Sun.

Beyond the stellar HZ, more massive planets have wider circumplanetary HZs. The HZ for a Mars-sized moon around $\beta$\,Pic\,b is between $18\,R_{\rm J}$ and $33\,R_{\rm J}$ for moon eccentricities $e\lesssim0.1$. As the planet ages, the HZ narrows and moves in. After a few hundred Myr, tidal heating may become the moon's major energy source. Owing to the strong dependence of tidal heating on the planet-moon distance, the HZ around $\beta$\,Pic\,b narrows to a few $R_{\rm J}$ within 1\,Gyr from now, details depending on $e$. An exomoon would have to reside in a very specific part of the $e$-$a$ space to be continuously habitable over a Gyr.

The high geometric transit probabilities of moons around giant planets, their higher transit frequencies, and the possibility of transit signals that are one order of magnitude deeper than that of the Earth around the Sun, make transit observations of moons around young giant planets a compelling science case for the upcoming \textit{GMT}, \textit{TMT}, and \textit{E-ELT} extremely large telescopes. Most intriguingly, these exomoons could orbit their young planets in the circumplanetary HZ and therefore be cradles of life.

\begin{acknowledgements}
This study has been inspired by discussions with Rory Barnes, to whom I express my sincere gratitude. The reports of an anonymous referee helped to clarify several passages of this paper. I have been supported by the German Academic Exchange Service, the Institute for Astrophysics G\"ottingen, and the Origins Institute at McMaster University, Canada. This work made use of NASA's ADS Bibliographic Services.
\end{acknowledgements}


\bibliographystyle{aa} 
\bibliography{ms}




\end{document}